\documentstyle[12pt]{article}
\newcommand{\bfp}{\mbox{\boldmath $p$}}

\newcommand{\bfk}{\mbox{\boldmath $k$}}
\def\nostrocostruttino#1\over#2{\mathrel{\mathop{\kern 0pt \rlap
{\hbox{$#1$}}} \hbox{\kern-.135em $#2$}}}
\def\sumint{\nostrocostruttino \sum \over {\displaystyle\int}}
\newcommand{\NP}[1]{{\it Nucl.\ Phys.}\ {\bf #1}}
\newcommand{\ZP}[1]{{\it Z.\ Phys.}\ {\bf #1}}
\newcommand{\PL}[1]{{\it Phys.\ Lett.}\ {\bf #1}}
\newcommand{\PR}[1]{{\it Phys.\ Rev.}\ {\bf #1}}
\newcommand{\PRL}[1]{{\it Phys.\ Rev.\ Lett.}\ {\bf #1}}

\def\lsim{\mathrel{\rlap{\lower4pt\hbox{\hskip1pt$\sim$}}\raise1pt\hbox{$<$}}} 
\def\gsim{\mathrel{\rlap{\lower4pt\hbox{\hskip1pt$\sim$}}\raise1pt\hbox{$>$}}} 
\begin{document}
%%%%%%%%%%%%%%%%%%%%%%%%%%%%%%%%%%%%%%%%%%%%%%%%%%%%%%%%%%%%%%%%%%%%%%%%%%%%%%
\begin{flushright}
DFTT 51/96 \\
INFNCA-TH9617 \\
hep-ph/9610407 \\
October 1996
\end{flushright}
\vskip 1.5cm
\begin{center}
{\bf 
Single spin asymmetries in DIS
}\\
\vskip 1.5cm
{\sf M.\ Anselmino$^1$, E. Leader$^2$ and F. Murgia$^3$}
\vskip 0.8cm
{$^1$Dipartimento di Fisica Teorica, Universit\`a di Torino and \\
      INFN, Sezione di Torino, Via P. Giuria 1, 10125 Torino, Italy\\
\vskip 0.5cm
 $^2$Birkbeck College, University of London, Malet Street,
      London WCIE 7HX, UK \\
\vskip 0.5cm
 $^3$INFN, Sezione di Cagliari, Via A. Negri 18, 09127 Cagliari, Italy } \\
\end{center}
\vskip 1.5cm
\noindent
{\bf Abstract:}

\vspace{6pt}

\noindent
We consider possible mechanisms for single spin asymmetries in inclusive  
Deep Inelastic Scattering (DIS) processes with unpolarized leptons and 
transversely polarized nucleons. Tests for the effects of non-zero intrinsic 
$\bfk_\perp$, for the properties of spin dependent quark fragmentations and  
for quark helicity conservation are suggested.
%%%%%%%%%%%%%%%%%%%%%%%%%%%%%%%%%%%%%%%%%%%%%%%%%%%%%%%%%%%%%%%%%%%%%%%%%%%%%%%
\newpage
%%%%%%%%%%%%%%%%%%%%%%%%%%%%%%%%%%%%%%%%%%%%%%%%%%%%%%%%%%%%%%%%%%%%%%%%%%%%%%%
\pagestyle{plain}
\setcounter{page}{1}

Single spin asymmetries in large $p_T$ inclusive hadronic reactions 
are forbidden in leading-twist perturbative QCD, reflecting the fact
that single spin asymmetries are zero at the partonic level and that
collinear parton configurations inside hadrons do not allow single 
spin dependences. Similarly, one might expect single spin asymmetries to 
vanish in large angle and high energy exclusive processes.
However, experiments tell us in several cases, both in inclusive
\cite{ada1,ada2} and exclusive reactions \cite{pol}, that single spin 
asymmetries are large and indeed non negligible.

The usual arguments to explain this apparent disagreement between pQCD 
and experiment invoke the moderate $p_T$ values of the data -- a few 
GeV, not quite yet in the true perturbative regime -- and the importance 
of higher-twist effects. Several phenomenological models have recently
attempted to explain the large single spin asymmetries observed in
$p^\uparrow p \to \pi X$ \cite{siv1}-\cite{noi}, as twist-3 effects which 
might be due to intrinsic partonic $\bfk_\perp$ in the fragmentation and/or 
distribution functions. Single spin effects in exclusive processes are harder 
to explain, as one cannot rely on the cross-section factorization theorem, 
as one does in the inclusive case, but has to deal with helicity amplitudes; 
in particular one needs quite significant single helicity flip partonic 
amplitudes which, however, are bound to be of ${\cal O}(\alpha_s m_q/\sqrt s)$ 
in pQCD, unless one resorts again to intrinsic $\bfk_\perp$ effects. 

We consider here a process in which one has convincing evidence that partons 
and perturbative QCD work well and successfully describe the unpolarized 
and leading-twist spin data, namely Deep Inelastic Scattering (DIS). In 
particular we shall discuss single spin asymmetries in the inclusive,
$\ell N^\uparrow \to \ell + jets$ and $\ell N^\uparrow \to hX$, 
reactions looking at possible origins of such asymmetries and 
devising strategies to isolate and discriminate among them.

According to the QCD hard scattering picture and the factorization theorem 
\cite{col1}-\cite{col3} the cross-section for the $\ell N^\uparrow \to hX$ 
reaction is given by

\begin{eqnarray}
& & \frac{E_h \, d^3\sigma^{\ell + N,S \to h + X}} {d^{3} \bfp_h} = 
\sum_{q; \lambda^{\,}_{\ell}, \lambda^{\,}_q, \lambda^\prime_q,
\lambda^{\,}_{q^{\prime}}, \lambda^{\prime}_{q^{\prime}}, \lambda^{\,}_h} 
{1 \over 2} \int \frac {dx \, d^2\bfk_\perp d^2\bfk_\perp^\prime} {\pi z} 
\frac {1}{16 \pi x^2 s^2} \label{gen} \\  
& & \rho^{q/N,S}_{\lambda^{\,}_q, \lambda^\prime_q}(x, \bfk_\perp) \,
\tilde f_{q/N}^{N,S}(x, \bfk_\perp) \>
\hat M^q_{\lambda^{\,}_{\ell}, \lambda^{\,}_{q^{\prime}}; 
\lambda^{\,}_{\ell}, \lambda^{\,}_q} \,
\hat M^{q \textstyle{*}}_{\lambda^{\,}_{\ell},
\lambda^{\prime}_{q^{\prime}}; 
\lambda^{\,}_{\ell}, \lambda^{\prime}_q} \>
\widetilde D_{\lambda^{\,}_h, \lambda^{\,}_h}^{
\lambda^{\,}_{q^{\prime}}, \lambda^{\prime}_{q^{\prime}}}
(z, \bfk_\perp^\prime) 
\nonumber
\end{eqnarray}
[wherever confusion is possible we label by $q^\prime$ the final 
quark, which is otherwise indicated by $q$].

Let us explain in some detail the meaning and physical content of the above 
equation. $\rho^{q/N,S}(x, \bfk_\perp)$ and 
$\tilde f_{q/N}^{N,S}(x, \bfk_\perp)$ are respectively the helicity density 
matrix and the total number density of quarks $q$  with momentum fraction 
$x$ and intrinsic transverse momentum $\bfk_\perp$ inside a polarized nucleon
$N$ with spin four-vector $S$. One can relate these quantities to the more 
familiar polarized structure functions; for example, for longitudinal 
polarization $S=S_L$ and in absence of intrinsinc transverse motion, one has 
\begin{equation}
\rho^{q/N,S_L}_{+,+}(x) \, f_{q/N}^{N,S_L}(x) = q_+(x) \,,
\end{equation}
where + stands for $\lambda^{\,}_q = 1/2$. In general 
$\rho^q_{\lambda^{\,}_q, \lambda^\prime_q}$ plays the same role as the 
density matrix of the initial state when describing a polarized 
scattering process \cite{bls}.

$\widetilde D_{\lambda^{\,}_h, \lambda^{\,}_h}^{\lambda^{\,}_q,
\lambda^{\prime}_q}(z, \bfk_\perp)$ describes the fragmentation process of 
a polarized quark $q$ into a hadron $h$ with helicity $\lambda^{\,}_h$, 
momentum fraction $z$ and intrinsic transverse momentum $\bfk_\perp$ with 
respect to the jet axis. It can be written in terms of the fragmentation
amplitudes for the $q \to h+X$ process as
\begin{equation}
\widetilde D_{\lambda^{\,}_h, \lambda^{\,}_h}^{\lambda^{\,}_q,
\lambda^{\prime}_q}(z, \bfk_\perp) = \sumint_{X,\lambda^{\,}_X} 
{\cal D}_{\lambda^{\,}_{X}, \lambda^{\,}_h; \lambda^{\,}_q}(z, \bfk_\perp) \,
{\cal D}^*_{\lambda^{\,}_{X}, \lambda^{\,}_h; \lambda^{\prime}_q}
(z, \bfk_\perp)
\label{fram}
\end{equation}
where the $\sumint_{X,\lambda^{\,}_X}$ stands for a spin 
sum and phase space integration of the undetected particles, considered as 
a system $X$. The usual unpolarized fragmentation function is simply
\begin{equation}
D_{h/q}(z) = {1\over 2} \sum_{\lambda^{\,}_q, \lambda^{\,}_h}
\int d^2\bfk_\perp \>
\widetilde D^{\lambda^{\,}_q,\lambda^{\,}_q}_{\lambda^{\,}_h,\lambda^{\,}_h}
(z, \bfk_\perp) \,.
\label{fr}
\end{equation}

Finally the $\hat M^{q}$s are the helicity amplitudes for the elementary 
lepton-quark reactions; they depend on $x, \bfk_\perp$ and 
$\bfk_\perp^\prime$ and their normalization is such that
\begin{equation}
{1 \over 2} \, \frac {1}{16 \pi x^2 s^2}
\sum_{\lambda^{\,}_{\ell}, \lambda^{\,}_q, \lambda^\prime_q}
\hat M^q_{\lambda^{\,}_{\ell}, \lambda^{\,}_{q^{\prime}}; 
\lambda^{\,}_{\ell}, \lambda^{\,}_q} \>
\hat M^{q \textstyle{*}}_{\lambda^{\,}_{\ell},
\lambda^{\prime}_{q^{\prime}}; \lambda^{\,}_{\ell}, \lambda^{\prime}_q} \>
\rho^{q/N,S}_{\lambda^{\,}_q, \lambda^\prime_q} =
\frac{d\hat\sigma^{q,P_q}}{d\hat t} \>
\rho^{q^{\prime}}_{\lambda^{\,}_{q^{\prime}}, \lambda^{\prime}_{q^{\prime}}}
\label{norm}
\end{equation}
where $d\hat\sigma^{q,P_q} / d\hat t$ is the cross-section for the $\ell 
q^\uparrow \to \ell q$ process, with an unpolarized lepton and an initial 
quark with polarization $P_q$ described by $\rho^{q/N,S}$, and 
$\rho^{q^{\prime}}_{\lambda^{\,}_{q^{\prime}}, \lambda^{\prime}_{q^{\prime}}}$ 
is the helicity density matrix of the final quark $q^{\prime}$ produced in 
such a process. Then Eq. (\ref{gen}) can be written in a more intuitive
way as 
\begin{eqnarray}
& & \frac{E_h \, d^3\sigma^{\ell + N,S \to h + X}} {d^{3} \bfp_h} = 
\sum_{q; \lambda^{\,}_{q^{\prime}}, \lambda^{\prime}_{q^{\prime}}, 
\lambda^{\,}_h} 
\int \frac {dx \, d^2\bfk_\perp d^2\bfk_\perp^\prime} {\pi z} \label{gen2} \\  
& & \tilde f_{q/N}^{N,S}(x, \bfk_\perp) \>
\frac{d\hat\sigma^{q,P_q}}{d\hat t}(x, \bfk_\perp, \bfk_\perp^\prime) \>
\rho^{q^{\prime}}_{\lambda^{\,}_{q^{\prime}}, \lambda^{\prime}_{q^{\prime}}}
(x, \bfk_\perp, \bfk_\perp^\prime) \> 
\widetilde D_{\lambda^{\,}_h, \lambda^{\,}_h}^{
\lambda^{\,}_{q^{\prime}}, \lambda^{\prime}_{q^{\prime}}}
(z, \bfk_\perp^\prime) \nonumber
\end{eqnarray}
where $\sum_{\lambda^{\,}_{q^{\prime}}, \lambda^{\prime}_{q^{\prime}}, 
\lambda^{\,}_h} \rho^{q^{\prime}}_{\lambda^{\,}_{q^{\prime}}, 
\lambda^{\prime}_{q^{\prime}}} \widetilde D_{\lambda^{\,}_h, 
\lambda^{\,}_h}^{\lambda^{\,}_{q^{\prime}}, \lambda^{\prime}_{q^{\prime}}}$
is simply the inclusive cross-section for the fragmentation process of the 
final polarized quark, $q^\prime \to h + X$. Such expressions are in general 
not diagonal in the helicity basis; 
in the case where the final quark is unpolarized 
$\rho^{q^{\prime}}_{\lambda^{\,}_{q^{\prime}}, \lambda^{\prime}_{q^{\prime}}}
= (1/2) \> \delta_{\lambda^{\,}_{q^{\prime}}, \lambda^{\prime}_{q^{\prime}}}$
and one recovers the usual expression for the unpolarized cross-section. 
Notice that for helicity conserving elementary interactions 
$d\hat\sigma^{q,P_q} / d\hat t$ equals the unpolarized cross-section 
$d\hat\sigma^q / d\hat t$.

Similar formulae hold also when the elementary interaction
is $\ell q \to \ell q g$ rather than $\ell q \to \ell q$: in the latter 
case two jets are observed in the final state -- the target jet and the 
current quark jet -- and in the former case three -- the target jet and 
$q$ + $g$ current jets.

In Eqs. (\ref{gen}) and (\ref{gen2}) we have taken into account possible 
intrinsic transverse momenta both in the distribution and the 
fragmentation process, together with a possible quark helicity non 
conservation in the elementary 
interactions ({\it e.g.}, $\lambda^{\,}_q \not= \lambda^{\,}_{q^{\prime}}$). 
Parity conservation allows, in general, non-zero single 
spin asymmetries under reversal of the nucleon spin, 
$d^3\sigma^{\ell + N,S \to h + X} \not= d^3\sigma^{\ell + N,-S \to h + X}$,
only for spin configurations with a non zero component perpendicular to 
the $\ell h$ production plane; a spin orientation perpendicular to 
such a plane would maximize the magnitude of the asymmetry. 

The $\bfk_\perp$ dependences are expected to have negligible effects on 
unpolarized variables for which they are indeed usually neglected, but 
they can be crucial for some single spin observables, as discussed in 
Refs. \cite{siv1}, \cite{siv2}, \cite{noi} and \cite{col1}; 
however, as a consequence of time reversal invariance, such effects cannot 
arise from the isolated process $p^\uparrow \to q + X$ (distribution function) 
or $q^\uparrow \to h + X$ (fragmentation function), but must involve some sort 
of initial state interactions between the proton and other particles in 
the reaction\footnote{The possibility of spin-isospin interactions has 
also been recently suggested \cite{ale}.}
or some final state interactions of the fragmenting quark. 
Such interactions are presumably always present in the case of fragmenting 
quarks; they are also expected, for the distribution functions, in some cases, 
{\it e.g.} in $pp$ interactions, but should be of higher order in 
$\alpha_{em}$ and therefore negligible in DIS.

In the case $\ell N^\uparrow \to hX$ with the observation of target + current
jets and eventually a final hadron inside a current jet one therefore remains 
with two possible theoretical sources of single spin asymmetries; in the quark 
fragmentation process and -- perhaps more unlikely, but not impossible -- in 
the elementary interactions. The former would confirm the suggestion of 
Collins \cite{col1}, whereas the latter would test much more fundamental 
properties of DIS, namely helicity conservation of the elementary QED and QCD 
hard interactions and the factorization theorem, which are usually taken 
for granted, but are still in need of definitive confirmation. 

We shall now describe a set of possible measurements which could shed light
on and test the above mechanisms. 

\vskip 6pt
\noindent
$a) \> \ell N^\uparrow \to \ell + 2\,jets$
\vskip 4pt

Here one avoids any fragmentation effect by looking at the fully 
inclusive cross-section for the process $\ell N^\uparrow \to \ell + 2\, jets$,
the 2 jets being the target and current ones; this is the usual DIS, 
the final quark spin is not observed, and one should set  
$\lambda^{\,}_{q^{\prime}}=\lambda^{\prime}_{q^{\prime}}$ so that 
Eq. (\ref{gen2}) becomes  
\begin{equation}
\frac{d^2\sigma^{\ell + N,S \to \ell + X}} {dx \, dQ^2} = \sum_q
\int d^2\bfk_\perp \> \tilde f_{q/N}^{N,S}(x, \bfk_\perp) \>
\frac{d\hat\sigma^{q,P_q}}{d\hat t}(x, \bfk_\perp) \,. 
\label{gena}
\end{equation}

In this case the elementary interaction is supposed to be a pure QED, 
helicity conserving one, $\ell q \to \ell q$, and $d\hat\sigma^{q,P_q}/d\hat t$
cannot depend on the quark polarization. Some spin dependence might 
remain in the distribution function, due to intrinsic $\bfk_\perp$ effects
\cite{siv1, siv2, noi}, but is expected to be of ${\cal O}(\alpha_{em})$.
The observation of a non-zero single spin asymmetry in such a process would 
quite seriously -- and utterly unexpectedly -- question the degree of 
validity of the one photon exchange approximation in DIS and the 
QCD factorization theorem, which takes into account soft and collinear
gluon emissions in the $Q^2$ dependent distribution functions.
 
\vskip 6pt
\noindent
$b) \> \ell N^\uparrow \to h + X \> (2\,jets, \> \bfk_\perp \not= 0)$
\vskip 4pt

One looks for a hadron $h$, with transverse momentum $\bfk_\perp$, inside 
the quark current jet; the final lepton may or may not be observed.
The elementary subprocess is $\ell q \to \ell q$ and Eq. (\ref{gen}) yields
\begin{eqnarray}
& & \frac{E_h \, d^5\sigma^{\ell + N,S \to h + X}} 
{d^{3} \bfp_h d^2 \bfk_\perp} = 
\sum_{q; \lambda^{\,}_{\ell}, \lambda^{\,}_q, \lambda^\prime_q, \lambda^{\,}_h} 
{1 \over 2} \int \frac {dx} {\pi z} 
\frac {1}{16 \pi x^2 s^2} \label{genk} \\  
&\times& \rho^{q/N,S}_{\lambda^{\,}_q, \lambda^\prime_q} \, f_{q/N}(x) \>
\hat M^q_{\lambda^{\,}_{\ell}, \lambda^{\,}_q; 
\lambda^{\,}_{\ell}, \lambda^{\,}_q} \,
\hat M^{q \textstyle{*}}_{\lambda^{\,}_{\ell}, \lambda^{\prime}_q; 
\lambda^{\,}_{\ell}, \lambda^{\prime}_q} \>
\tilde D_{\lambda^{\,}_h, \lambda^{\,}_h}^{
\lambda^{\,}_q, \lambda^{\prime}_q}
(z, \bfk_\perp) \,.
\nonumber
\end{eqnarray}
where we have neglected intrinsic $\bfk_\perp$ effects in the distribution
functions, as they are expected to be of ${\cal O}(\alpha_{em})$.
Eq. (\ref{genk}) is diagonal in the transverse spin basis and 
leads to the single spin asymmetry for transversely polarized nucleons:
\begin{eqnarray}
& & \frac{E_h \, d^5\sigma^{\ell + N^\uparrow \to h + X}} 
{d^{3} \bfp_h d^2 \bfk_\perp} 
- \frac{E_h \, d^5\sigma^{\ell + N^\downarrow \to h + X}} 
{d^{3} \bfp_h d^2 \bfk_\perp} \label{coll} \\
&=& \sum_q \int \frac {dx} {\pi z} \>    
\Delta_{_T} q(x) \> \Delta_{_N} \hat\sigma^q (x, \bfk_\perp) \,
\left[ \tilde D_{h/q^\uparrow}(z, \bfk_\perp)
- \tilde D_{h/q^\uparrow}(z, - \bfk_\perp) \right]
\nonumber
\end{eqnarray}
where $\Delta_{_T}q$ is the polarized number density for transversely spinning 
quarks $q$ and $\Delta_{_N} \hat\sigma^q$ is the elementary cross-section 
double spin asymmetry
\begin{equation}
\Delta_{_N} \hat\sigma^q = {d\hat \sigma^{\ell q^\uparrow \to 
\ell q^\uparrow} \over d\hat t} - {d\hat \sigma^{\ell q^\uparrow \to 
\ell q^\downarrow} \over d\hat t} \,\cdot
\label{del}
\end{equation}

The quantity in square brackets on the r.h.s. of Eq. (\ref{coll}) could 
be non zero \cite{col1} and a measurement of the l.h.s. would be a clear 
test of such suggestion. Notice that even upon integration over
$d^2\bfk_\perp$ the spin asymmetry of Eq. (\ref{coll}) might survive,
due to some $\bfk_\perp$ dependence in $\Delta_{_N} \hat\sigma^q$: the 
original leading-twist Collins effect in the fragmentation will be 
diminished by $k_\perp/p_T$ higher twist terms and there might be 
cancellations between different quark contributions, but some overall 
effect might remain if one considers fast particles inside the current
jets, so that only valence polarized quarks from the polarized nucleon
contribute. 

\vskip 6pt
\noindent
$c) \> \ell N^\uparrow \to h + X \> (2\,jets, \> \bfk_\perp = 0)$
\vskip 4pt

By selecting events with the final hadron collinear to the jet axis
($\bfk_\perp = 0$) one forbids any single spin effect in the fragmentation
process. As in the fully inclusive case $a)$ the observation of a
single spin asymmetry in such a case would require reconsideration
of the degree of validity of the QED helicity conserving one photon exchange 
dominance and of QCD factorization theorem in DIS.

\vskip 6pt
\noindent
$d) \> \ell N^\uparrow \to h + X \> (3\,jets, \> \bfk_\perp \not= 0)$
\vskip 4pt
 
The elementary process is now $\ell q \to \ell qg$ and one looks at 
hadrons with $\bfk_\perp \not= 0$ inside the $q$ current jet. Single
spin asymmetries can originate from the Collins effect in the 
fragmentation process, analogously to what was discussed in point $b)$ .

\vskip 6pt
\noindent
$e) \> \ell N^\uparrow \to \ell + 3\,jets$ or 
$\ell N^\uparrow \to h + X \> (3\,jets, \> \bfk_\perp = 0)$
\vskip 4pt

These cases are analogous to $a)$ and $c)$ respectively: the measurement 
eliminates spin effects arising from the distribution and fragmentation 
functions. The only possible origin of a single spin asymmetry would reside 
in the elementary interaction, which is now a hard perturbative QCD process, 
$\ell q \to \ell qg$. Single spin asymmetries require single quark 
helicity flip and the observation of such an asymmetry in this case would 
question quark helicity conservation, a fundamental property of pQCD which 
has never been directly tested. 

In summary, a study of single transverse spin asymmetries in DIS could
provide a series of profound tests of our understanding of large $p_T$
QCD-controlled reactions\footnote{Upon completion of this work we became 
aware of a somewhat similar analysis \cite{men2}.}.

\vskip 12pt
\noindent
{\bf Acknowledgements}. E.L. acknowledges the support of the Istituto
Nazionale di Fisica Nucleare and the Research Committee of Birkbeck College.

\end{document}